\documentclass[11pt,a4paper]{article}

\usepackage[utf8]{inputenc}
\usepackage[T1]{fontenc}
\usepackage{lmodern}
\usepackage[margin=1in]{geometry}
\usepackage{amsmath,amssymb}
\usepackage{booktabs}
\usepackage{graphicx}
\usepackage{xcolor}
\usepackage{hyperref}
\usepackage{cleveref}
\usepackage{enumitem}
\usepackage{float}
\usepackage{caption}
\usepackage{subcaption}
\usepackage{pgfplots}
\usepackage{pgfplotstable}
\pgfplotsset{compat=1.18}
\graphicspath{{figures/}}
\usepackage{natbib}
\usepackage{microtype}
\usepackage{tabularx}
\usepackage{multirow}

\definecolor{critical}{HTML}{D32F2F}
\definecolor{alarming}{HTML}{E64A19}
\definecolor{concerning}{HTML}{F9A825}
\definecolor{notable}{HTML}{1976D2}
\definecolor{curious}{HTML}{388E3C}

\title{\textbf{Arbiter: Detecting Interference in LLM Agent System Prompts}\\[0.5em]
\large A Cross-Vendor Analysis of Architectural Failure Modes}

\author{Tony Mason \\ the University of British Columbia \\ the Georgia Institute of Technology \\ fsgeek@\{cs.ubc.ca,gatech.edu,wamason.com\} }

\date{March 2026}

\begin{document}

\maketitle

\noindent\textbf{Note on version 2 (October 2026).}\enspace
This version corrects three statements about the paper's own method.
The measurements of the February 2026 prompts are unchanged. (1)~The
three-consecutive-``no'' stopping criterion was applied in full only to
the Claude Code campaign; the Codex and Gemini campaigns stopped at the
first declining vote. (2)~Version~1 named human review of all scourer
findings and cross-model confirmation as mitigations for LLM
fabrication. No per-finding human verdicts were recorded, and the
map-passing protocol instructs each model not to repeat prior findings,
which rules out the confirmation that was claimed. (3)~The claim that
multi-model evaluation discovers categorically different vulnerability
classes is restated as a hypothesis: the protocol's non-repetition
instruction confounds it, and the same-model control that would
separate the two was not run. These corrections follow an external
review (October 2026) and the author's audit of the stored campaign
data; the stopping votes and per-pass models are in the repository's
\texttt{data/scourer/} files.

\begin{abstract}
System prompts for LLM-based coding agents are software artifacts that
govern agent behavior, yet lack the testing infrastructure applied to
conventional software. We present Arbiter, a framework combining formal
evaluation rules with multi-model LLM scouring to detect interference
patterns in system prompts. Applied to three major coding agent system
prompts---Claude Code (Anthropic), Codex CLI (OpenAI), and Gemini CLI
(Google)---we identify 152 findings across the undirected scouring
phase and 21 hand-labeled interference patterns in directed analysis
of one vendor. We show that prompt architecture (monolithic, flat,
modular) strongly correlates with observed \emph{failure class} but not
with
\emph{severity}, and that successive models in a map-passing scouring
chain surface different categories of findings---an observation our
design cannot separate from the protocol's instruction not to repeat
prior findings. One scourer finding---structural data loss in Gemini CLI's
memory system---was independently confirmed when Google filed and
patched the symptom without addressing the schema-level root cause
identified by the scourer. Total cost of cross-vendor analysis: \$0.27
USD.
\end{abstract}

\section{Introduction}
\label{sec:introduction}

LLM-based coding agents---Claude Code, Codex CLI, Gemini CLI, and
others---are governed by system prompts ranging from 245 to 1,490
lines. These prompts are software artifacts in every meaningful sense:
they specify behavior, encode precedence hierarchies, define tool
contracts, manage state, and compose subsystems. A system prompt is the
constitution under which the agent operates.

In this paper, we also use a scoped ``prompts as contracts'' framing:
contract means precedence-bound interpretation with explicit conflict
adjudication. It does \emph{not} imply bilateral negotiation, legal
remedy, or meeting-of-the-minds assumptions. The value of the framing is
operational: it clarifies standing (which component adjudicates), canon
application (how ambiguities are interpreted), and refusal conditions
when instructions are mutually unsatisfiable.

Unlike conventional software, these constitutions have no type checker,
no linter, no test suite. When a system prompt contains contradictory
instructions---``always use TodoWrite'' in one section, ``NEVER use
TodoWrite'' in another---the executing LLM resolves the conflict
silently through whatever heuristic its training provides. No error is
raised. No warning is logged. The contradiction persists, and the
agent's behavior becomes a function of which instruction the model
happens to weight more heavily on a given invocation.

This paper argues that \textbf{the agent that resolves the conflict
cannot be the agent that detects it.} An LLM executing a contradictory
system prompt will smooth over inconsistencies through
``judgment''---the same mechanism that makes LLMs useful makes them
unreliable as their own auditors. External evaluation against formal
criteria is required.

We present Arbiter, a framework with two complementary evaluation
phases:

\begin{enumerate}[leftmargin=*]
  \item \textbf{Directed evaluation} decomposes a system prompt into
    classified blocks and evaluates block pairs against formal
    interference rules. This is exhaustive within its defined search
    frame: if a rule exists for a failure mode, and two blocks share
    the relevant scope, the evaluation is guaranteed to check that pair.
  \item \textbf{Undirected scouring} sends the prompt to multiple LLMs
    with deliberately vague instructions---``read this carefully and
    note what you find interesting.'' Each pass receives the accumulated
    findings from all prior passes and is asked to go where previous
    explorers didn't. Convergent termination (three consecutive models
    declining to continue) was applied to the Claude Code campaign; the
    Codex and Gemini campaigns stopped at the first declining vote.
\end{enumerate}

Applied to three major coding agent system prompts, these two phases
reveal 152 scourer findings and 21 hand-labeled interference patterns.
The findings organize into a taxonomy correlated with prompt
architecture: monolithic prompts produce growth-level bugs at subsystem
boundaries, flat prompts trade capability for consistency, and modular
prompts produce design-level bugs at composition seams.

Our contributions are:
\begin{itemize}[leftmargin=*]
  \item A taxonomy of system prompt failure modes grounded in
    cross-vendor empirical evidence.
  \item A methodology for systematic prompt analysis combining directed
    rules with undirected multi-model scouring.
  \item The observation that successive models in a map-passing chain
    surface different categories of findings, stated as a hypothesis
    about model complementarity that the present design does not test
    (no same-model chain was run).
  \item External validation: a scourer-identified design bug
    independently confirmed by the vendor's own bug report and patch.
  \item Demonstration that comprehensive cross-vendor analysis costs
    \$0.27 in API calls---less than three minutes of US minimum wage
    labor.
\end{itemize}

\section{Background and Related Work}
\label{sec:background}

\subsection{Prompt Engineering}

The prompt engineering literature has grown rapidly since 2023, but
overwhelmingly focuses on crafting prompts for specific tasks: few-shot
formatting, chain-of-thought elicitation~\citep{wei2022chain},
tree-of-thoughts~\citep{yao2023tree}, retrieval-augmented generation
patterns. The quality of the prompt itself---as a software artifact with
internal consistency requirements---receives comparatively little
attention. \citet{gloaguen2026evaluating} provide the closest
empirical study: they evaluate repository-level context files
(AGENTS.md) and find that LLM-generated instructions \emph{reduce}
agent performance while increasing cost, concluding that ``unnecessary
requirements make tasks harder.'' Their work measures runtime impact of
instruction quality; ours identifies the structural conditions that
produce those unnecessary requirements.

\subsection{Prompt Security}

Prompt injection research~\citep{schulhoff2023ignore, greshake2023not,
liu2023prompt} examines adversarial inputs designed to override system
prompt instructions. This work is orthogonal to ours: we analyze the
system prompt itself for internal contradictions, not external attacks
against it. A system prompt can be perfectly secure against injection
yet internally contradictory.

\subsection{Software Architecture Failure Modes}

The analogy between system prompt architectures and software
architectures is deliberate. Monolithic applications accumulate
contradictions at subsystem boundaries as teams add features
independently~\citep{conway1968committees}. Microservice architectures
push bugs to composition seams---each service works internally, but
contracts between services may be
inconsistent~\citep{newman2021building}. The module decomposition
criteria identified by~\citet{parnas1972criteria} apply directly:
information hiding, interface specification, and the consequences of
design decisions that cross module boundaries. These same patterns
appear in system prompts, and for the same structural reasons.

\subsection{LLM-as-Judge}

Using LLMs to evaluate LLM outputs is established
practice~\citep{zheng2023judging}. We extend this to multi-model
evaluation, where the goal is not consensus but complementarity:
different models, trained on different data, bring different analytical
biases. Self-consistency~\citep{wang2022self} demonstrates that
sampling diverse reasoning paths improves single-model reliability; we
apply the same principle across models, where diversity comes from
training rather than sampling.

\section{Methodology}
\label{sec:methodology}

\subsection{Corpus}
\label{sec:corpus}

We analyze three publicly available system prompts from major coding
agent vendors:

\begin{table}[H]
\centering
\caption{System prompt corpus.}
\label{tab:corpus}
\begin{tabular}{@{}llrrl@{}}
\toprule
Agent & Vendor & Lines & Chars & Source \\
\midrule
Claude Code & Anthropic & 1{,}490 & 78K & npm package \\
Codex CLI & OpenAI & 298 & 22K & open-source repo \\
Gemini CLI & Google & 245 & 27K & TS render functions \\
\bottomrule
\end{tabular}
\end{table}

All three prompts are derived from publicly available artifacts. Claude
Code's system prompt was extracted from the published npm package.
Codex CLI's prompt exists as a plaintext Markdown file in the
open-source repository. Gemini CLI's prompt required composition: the
open-source repository contains TypeScript render functions
(\texttt{renderPreamble()}, \texttt{renderCoreMandates()}, etc.)
assembled at runtime with feature flags controlling which sections are
included. We wrote a renderer to compose all sections with maximal
feature flags enabled, producing a 245-line prompt representing the
largest attack surface.

The prompts span an order-of-magnitude size range. Claude Code's prompt
is 5$\times$ longer than Codex's and 6$\times$ longer than Gemini
CLI's. This variation reflects fundamentally different architectural
choices about how much to encode in the system prompt versus in tool
definitions, model training, or runtime logic.

\subsection{Directed Evaluation: Prompt Archaeology}
\label{sec:directed}

The directed phase proceeds in three steps: decomposition, rule
application, and interference pattern identification.

\paragraph{Decomposition.} The prompt is broken into contiguous blocks,
each classified by tier (system/domain/application), category
(identity, security, tool-usage, workflow, etc.), modality (mandate,
prohibition, guidance, information), and scope (what topics or tools
the block governs). For Claude Code v2.1.50, this produced 56
classified blocks.

\paragraph{Rule application.} Formal evaluation rules define the
interference types to check for. Five built-in rules derive from the
initial archaeology:

\begin{table}[H]
\centering
\caption{Built-in evaluation rules.}
\label{tab:rules}
\begin{tabular}{@{}lll@{}}
\toprule
Rule & Type & Detection \\
\midrule
Mandate-prohibition conflict & Direct contradiction & LLM + structural \\
Scope overlap redundancy & Scope overlap & LLM \\
Priority marker ambiguity & Priority ambiguity & Structural \\
Implicit dependency & Unresolved dependency & LLM \\
Verbatim duplication & Scope overlap & Structural \\
\bottomrule
\end{tabular}
\end{table}

Structural rules (priority markers, verbatim duplication) run as Python
predicates---no LLM call, no cost. LLM rules use per-rule prompt
templates evaluated against block pairs.

\paragraph{Pre-filtering.} Not all $O(n^2 \times R)$ block-pair-rule
combinations are evaluated. Rules specify pre-filters: scope overlap
requirements, modality constraints. For 56 blocks and 5 rules,
pre-filtering reduces the search space from approximately 15{,}680
evaluations to 100--200 relevant pairs.

\paragraph{Results on Claude Code.} Directed analysis of Claude Code
v2.1.50 identified 21 interference patterns:

\begin{itemize}[leftmargin=*]
  \item 4 \textbf{critical} direct contradictions: TodoWrite mandate
    (``use VERY frequently,'' ``ALWAYS use'') directly conflicts with
    commit and PR workflow prohibitions (``NEVER use TodoWrite''). A
    model in a commit context must violate one instruction or the other.
  \item 13 \textbf{scope overlaps}: the same behavioral constraint
    (read-before-edit, no-emoji, prefer-dedicated-tools, no-new-files)
    restated in 2--3 locations with subtle differences.
  \item 2 \textbf{priority ambiguities}: parallel execution guidance
    coexists with commit workflow's specific sequential ordering;
    security policy appears verbatim at two locations.
  \item 2 \textbf{implicit dependencies}: plan mode restricts
    file-editing tools while general policy prohibits using Bash for
    file operations, creating an undeclared dead zone.
\end{itemize}

Of these 21 patterns, 20 (95\%) were statically detectable---a
compiler checking scope overlap and modality conflict would catch them
without any LLM involvement. The one non-static pattern was the
plan-mode dead zone (implicit dependency), which requires semantic
reasoning over cross-section interactions rather than local syntax.

\subsection{Undirected Evaluation: Multi-Model Scouring}
\label{sec:scouring}

The directed phase is exhaustive within its search frame but blind
outside it. A rule for ``mandate-prohibition conflict'' will find every
instance, but it cannot find vulnerability classes for which no rule
exists. The undirected scouring phase addresses this gap.

\paragraph{Design.} A scourer prompt is deliberately vague:

\begin{quote}
\itshape
You are exploring a system prompt. Not auditing it, not checking it
against rules---just reading it carefully and noting what you find
interesting. ``Interesting'' is deliberately vague. Trust your judgment.
\end{quote}

The scourer asks the LLM to classify its own findings (freeform
categories) and rate their severity on a four-level epistemic scale:
\emph{curious} (pattern noticed), \emph{notable} (worth
investigating), \emph{concerning} (likely problematic), \emph{alarming}
(structurally guaranteed to cause failures).

\paragraph{Multi-pass composition.} Each scourer pass receives the
complete findings and unexplored territory from all prior passes. This
map-passing composition ensures that later passes explore genuinely new
territory rather than rediscovering what earlier passes found.

\paragraph{Multi-model execution.} Each pass uses a different LLM.
This is the key methodological choice: different models bring different
analytical biases rooted in their training data and architecture. The
goal is not consensus but complementarity---we want models that
disagree about what's interesting, because disagreement reveals the
space of possible analyses.

\paragraph{Convergent termination.} Each pass independently assesses
whether another pass would find new material. When three consecutive
models decline to continue (setting \texttt{should\_send\_another:
false}), we treat the exploration as converged. This rule was applied
in full only to the Claude Code campaign, where passes 8--10 declined.
The Codex and Gemini campaigns were stopped at the first declining vote
(pass~2 and pass~3 respectively). Version~1 of this paper stated that
the criterion transferred across all three prompts without
recalibration; that overstated what was run.

\subsection{Scourer Execution}
\label{sec:scourer-execution}

\begin{table}[H]
\centering
\caption{Scourer campaign summary.}
\label{tab:scourer-summary}
\begin{tabular}{@{}lrrrl@{}}
\toprule
Vendor & Passes & Models & Findings & Stopping Signal \\
\midrule
Claude Code & 10 & 10 & 116 & 3 consecutive ``no'' \\
Codex CLI & 2 & 2 & 15 & Pass 2 said ``enough'' \\
Gemini CLI & 3 & 3 & 21 & Pass 3 said ``enough'' \\
\bottomrule
\end{tabular}
\end{table}

\textbf{Models used across all analyses:} Claude Opus 4.6, Gemini 2.0
Flash, Kimi K2.5, DeepSeek V3.2, Grok 4.1, Llama 4 Maverick, MiniMax
M2.5, Qwen3-235B, GLM 4.7, GPT-OSS 120B.

Claude Code required 10 passes to converge---consistent with its
5--6$\times$ larger size. The smaller prompts converged in 2--3 passes.
This suggests a roughly logarithmic relationship between prompt size
and passes to convergence, though three data points cannot confirm
this.

\subsection{Complementarity of Phases}
\label{sec:complementarity}

The directed and undirected phases are not competing approaches. They
are two phases of the same analysis, each finding what the other
cannot:

\textbf{Directed rules find exhaustive enumerations.} If a rule exists
for scope overlap, the directed phase will find every instance---all 13
in Claude Code. No scourer pass will systematically enumerate all
pairs.

\textbf{Undirected scouring finds what's outside the search frame.}
The scourer discovered vulnerability classes for which no rule existed:
security architecture gaps (system-reminder trust as injection
surface), economic exploitation vectors (unbounded token generation),
operational risks (context compression deleting saved preferences),
identity confusion, and implementation language leaks. These categories
were invisible to directed analysis because no rule had been written
for them.

The pattern is familiar from software engineering: static analysis
catches what unit tests miss, and vice versa. Neither subsumes the
other. The value is in running both.

\subsection{Structural Analysis: Prompt AST}
\label{sec:ast}

Both the directed and undirected phases analyze what the prompt
\emph{says}. A third analytical layer examines what the prompt
\emph{is}---its structural properties as a document.

We built a two-layer parser that produces an abstract syntax tree (AST)
for any system prompt:

\paragraph{Structural layer.} The parser decomposes a prompt into a
tree of typed nodes: Document $\to$ Section $\to$ (Paragraph $|$
Directive $|$ List $|$ CodeBlock $|$ Metadata). This is
vendor-neutral---it operates on the document's physical structure
(headings, bullet lists, code fences), not on any vendor's conventions.

\paragraph{Semantic layer.} Each node is classified by semantic role
(identity, policy, safety, tool\_usage, workflow, format,
memory\_policy, environment, meta) and assigned to a channel (behavior,
tool\_schema, memory, environment). Role assignment uses the node's
content and its position in the section hierarchy---a ``NEVER'' under a
Safety heading carries different semantics than a ``NEVER'' under Tone
and Style.

\paragraph{Structural hashing.} Each node receives a
content-independent structural hash based on its type, depth, and
position among siblings. This enables clone detection across prompt
versions: two nodes with the same structural hash occupy the same
position in the document skeleton, regardless of whether their content
has changed.

\paragraph{AST diffing.} Given two ASTs, the differ classifies every
node as added, removed, modified (same structural hash, different
content), or moved (same content hash, different structural hash). This
produces a precise changelog between prompt versions without relying on
line-level text diffing, which is confounded by reformatting.

\section{Results}
\label{sec:results}

\subsection{Quantitative Summary}
\label{sec:quantitative}

\begin{table}[H]
\centering
\caption{Cross-vendor analysis summary.}
\label{tab:results-summary}
\begin{tabular}{@{}lrrr@{}}
\toprule
Metric & Claude Code & Codex CLI & Gemini CLI \\
\midrule
Lines & 1{,}490 & 298 & 245 \\
Characters & 78K & 22K & 27K \\
Scourer findings & 116 & 15 & 21 \\
Passes to convergence & 10 & 2 & 3 \\
Distinct models & 10 & 2 & 3 \\
Archaeology patterns & 21 & --- & --- \\
Worst (scourer) & alarming (12) & concerning (5) & alarming (2) \\
Worst (archaeology) & critical (4) & --- & --- \\
Actual API cost & \$0.236 & \$0.012 & \$0.014 \\
\bottomrule
\end{tabular}
\end{table}

\textbf{Total API cost across all three analyses: \$0.27} (OpenRouter
billing data; Claude Opus pass billed via subscription, excluded).

Directed archaeology was performed only on Claude Code, where 56
hand-labeled blocks and 21 hand-labeled interference patterns serve as
ground truth. The scourer was run on all three vendors.

\subsection{Severity Distributions}
\label{sec:severity}

Scourer severity uses a four-level epistemic scale:

\begin{table}[H]
\centering
\caption{Severity distribution across vendors.}
\label{tab:severity}
\begin{tabular}{@{}lrrr@{}}
\toprule
Severity & Claude Code & Codex CLI & Gemini CLI \\
\midrule
Curious & 34 (29\%) & 3 (20\%) & 4 (19\%) \\
Notable & 36 (31\%) & 7 (47\%) & 9 (43\%) \\
Concerning & 34 (29\%) & 5 (33\%) & 6 (29\%) \\
Alarming & 12 (10\%) & 0 (0\%) & 2 (10\%) \\
\bottomrule
\end{tabular}
\end{table}

Claude Code's distribution is notably uniform across the lower three
levels, with a long tail of alarming findings. Codex and Gemini CLI
both peak at ``notable''---more findings are interesting-but-not-dangerous
than dangerous. This likely reflects Claude Code's size: a 1{,}490-line
prompt has more surface area for serious contradictions.

\begin{figure}[H]
\centering
\includegraphics[width=0.8\textwidth]{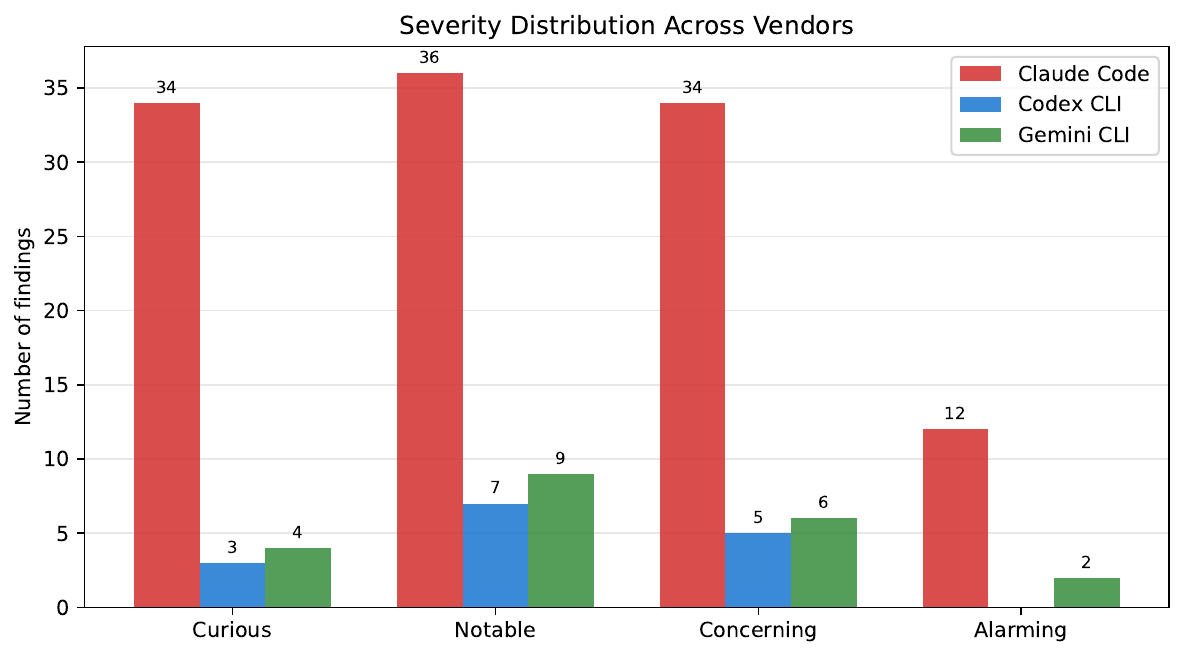}
\caption{Severity distribution across vendors. Claude Code shows a
  nearly uniform distribution across the lower three levels with an
  alarming tail. Smaller prompts (Codex, Gemini) peak at ``notable.''}
\label{fig:severity}
\end{figure}

\subsection{Architecture-Determined Failure Modes}
\label{sec:architecture}

The central finding of this analysis is that prompt architecture
strongly correlates with observed failure mode class. Three
architectural patterns produce three
characteristic classes of bug.

\subsubsection{Monolith: Claude Code}

Claude Code's system prompt is a single 1{,}490-line document
containing identity declarations, security policy, behavioral
guidelines, 15+ tool definitions with usage instructions, workflow
templates (commit, PR, planning), a task management system, team
coordination protocol, and agent spawning infrastructure. It grew by
accretion: subsystems were developed independently and composed into a
single artifact.

\textbf{Characteristic failure mode: growth-level bugs at subsystem
boundaries.} The four critical contradictions all follow the same
pattern: a general-purpose subsystem (TodoWrite task management) makes
universal claims (``ALWAYS use,'' ``use VERY frequently'') that
conflict with specific workflow subsystems (commit, PR creation) that
prohibit the same tool (``NEVER use TodoWrite''). These subsystems
were evidently authored by different teams or at different times, and
no integration test checks their mutual consistency.

The 13 scope overlaps follow the same structural logic. The
read-before-edit constraint appears in three places: a general
behavioral section, the Edit tool definition, and the Write tool
definition. Each instance is worded slightly differently. Today they're
consistent; tomorrow, when someone edits one without updating the
others, they won't be.

\textbf{Monolith prognosis:} These bugs are fixable by
refactoring---extract the duplicated constraints into single-source
declarations, scope the TodoWrite mandate to exclude workflow-specific
contexts. The architecture itself isn't wrong, but it needs the same
maintenance discipline that monolithic codebases require.

\subsubsection{Flat: Codex CLI}

Codex CLI's system prompt is 298 lines---the shortest of the three and
the simplest in structure. No nested prompts, no team coordination, no
agent spawning.

\textbf{Characteristic failure mode: simplicity trade-offs.} With fewer
capabilities encoded in the prompt, there are fewer opportunities for
contradiction. The 15 findings are overwhelmingly structural
observations rather than operational bugs: identity confusion between
``GPT-5.2'' (the model) and ``Codex CLI'' (the tool), sequential plan
status tracking versus parallel tool execution, an empty
``Responsiveness'' section header suggesting truncation, and leaked
implementation details (explicit suppression of an inline citation
format).

\textbf{Flat prognosis:} Codex's simplicity buys consistency. The
prompt is the cleanest of the three.

\subsubsection{Modular: Gemini CLI}

Gemini CLI's system prompt is 245 lines, composed from modular
TypeScript render functions assembled at runtime with feature flags.

\textbf{Characteristic failure mode: design-level bugs at composition
seams.} Each module works correctly in isolation. The bugs exist
exclusively in the gaps between modules---contracts that were never
specified because each module was designed independently.

Two findings rated ``alarming'' by independent scourer models:

\paragraph{Structural data loss during history compression.} Gemini CLI
includes a nested ``History Compression System Prompt'' that governs
how conversation history is summarized when context limits are reached.
This compression prompt defines a rigid XML schema
(\texttt{<state\_snapshot>}) that becomes ``the agent's only memory''
post-compression. The \texttt{save\_memory} tool allows users to store
global preferences. However, the compression schema contains no field
for saved memories. Consequently, any preference stored via
\texttt{save\_memory} is \textbf{structurally guaranteed to be deleted}
during a compression event.

This is not a bug in either module. The \texttt{save\_memory} tool
works correctly. The compression prompt works correctly. The bug exists
in the contract between them---a contract that was never written.

\textbf{Post-hoc validation.} After the scourer identified this
finding, we discovered that Google had independently filed a P0 bug
(\texttt{google-gemini/gemini-cli\#16213}) about the compression system
entering an infinite loop. The merged fix (PR \#16914, +748/$-$56
lines, January 2026) reorders the compression pipeline and adds token
budget truncation. It makes compression \emph{work}. It does not add a
\texttt{<saved\_memory>} field to the compression schema, does not
change how \texttt{save\_memory} data flows through compression, and
does not mention user preference persistence in the code review
discussion. The fix resolves the symptom while leaving the schema-level
data loss intact.

\paragraph{Impossible simultaneous compliance.} The ``Explain Before
Acting'' mandate requires a one-sentence explanation before every tool
call. The ``Minimal Output'' rule mandates output of fewer than three
lines and prohibits ``mechanical tool-use narration.'' In any
non-trivial task requiring multiple tool calls, simultaneous compliance
is structurally impossible.

\textbf{Modular prognosis:} These bugs cannot be fixed by editing one
module. They require changing the architectural contracts between
modules.

\subsection{Universal Patterns}
\label{sec:universal}

Three interference patterns appear in all three system prompts,
suggesting they are inherent to the task of governing an LLM coding
agent:

\textbf{Autonomy versus restraint.} All three prompts contain
instructions to ``persist until the task is fully handled'' alongside
instructions to ``ask before acting'' on ambiguous tasks. The tension
is inherent: a useful coding agent must be both autonomous enough to
complete tasks and cautious enough not to cause damage.

\textbf{Precedence hierarchy ambiguity.} All three prompts define
multiple authority sources---system instructions, configuration files,
user messages, tool-injected context---without fully specifying how
conflicts between them resolve.

\textbf{State-dependent behavioral modes.} All three prompts include
mechanisms for changing the agent's behavior based on runtime
state---approval presets, plan mode, skill activation---that change
which rules apply without always specifying how mode-specific rules
interact with base rules.

\subsection{Multi-Model Complementarity}
\label{sec:complementarity-results}

The most methodologically significant finding concerns what different
models discover. Each model brings analytical biases rooted in its
training:

\begin{table}[H]
\centering
\caption{Per-model analytical focus areas.}
\label{tab:model-focus}
\begin{tabular}{@{}lp{8cm}@{}}
\toprule
Model & Characteristic Focus \\
\midrule
Claude Opus 4.6 & Structural contradictions, security surfaces, meta-observations \\
DeepSeek V3.2 & Hidden references, delegation loopholes, format mismatches \\
Kimi K2.5 & Economic exploitation, resource exhaustion, cognitive load \\
Grok 4.1 & Permission schema gaps, environment assumptions, state management \\
Llama 4 Maverick & Constraint inconsistency, security boundaries \\
MiniMax M2.5 & Trust architecture flaws, concurrency, impossible instructions \\
Qwen3-235B & Contextual contradictions, state preservation illusions \\
GLM 4.7 & Data integrity, temporal paradoxes, competitive logic \\
\bottomrule
\end{tabular}
\end{table}

This is not ``more models find more findings.'' It is \textbf{different
models find different kinds of findings.} Kimi K2.5's
economic/resource lens is categorically absent from Claude Opus's
structural analysis. GLM's focus on data integrity does not overlap
with Grok's attention to permission schemas. The models are not
redundant; they are complementary.

This observation is confounded with the protocol. Every pass receives
all prior findings and is instructed to go where previous explorers
did not, so divergence between successive models is what the
instruction demands of any model. The design that would separate model
bias from instruction---the same model run for $N$ passes under the
same instruction, and the model order permuted---was not run for
version~1. Until it is, the per-model focus areas in
\Cref{tab:model-focus} describe one chain, not model-specific
analytical bias.

\begin{figure}[H]
\centering
\includegraphics[width=\textwidth]{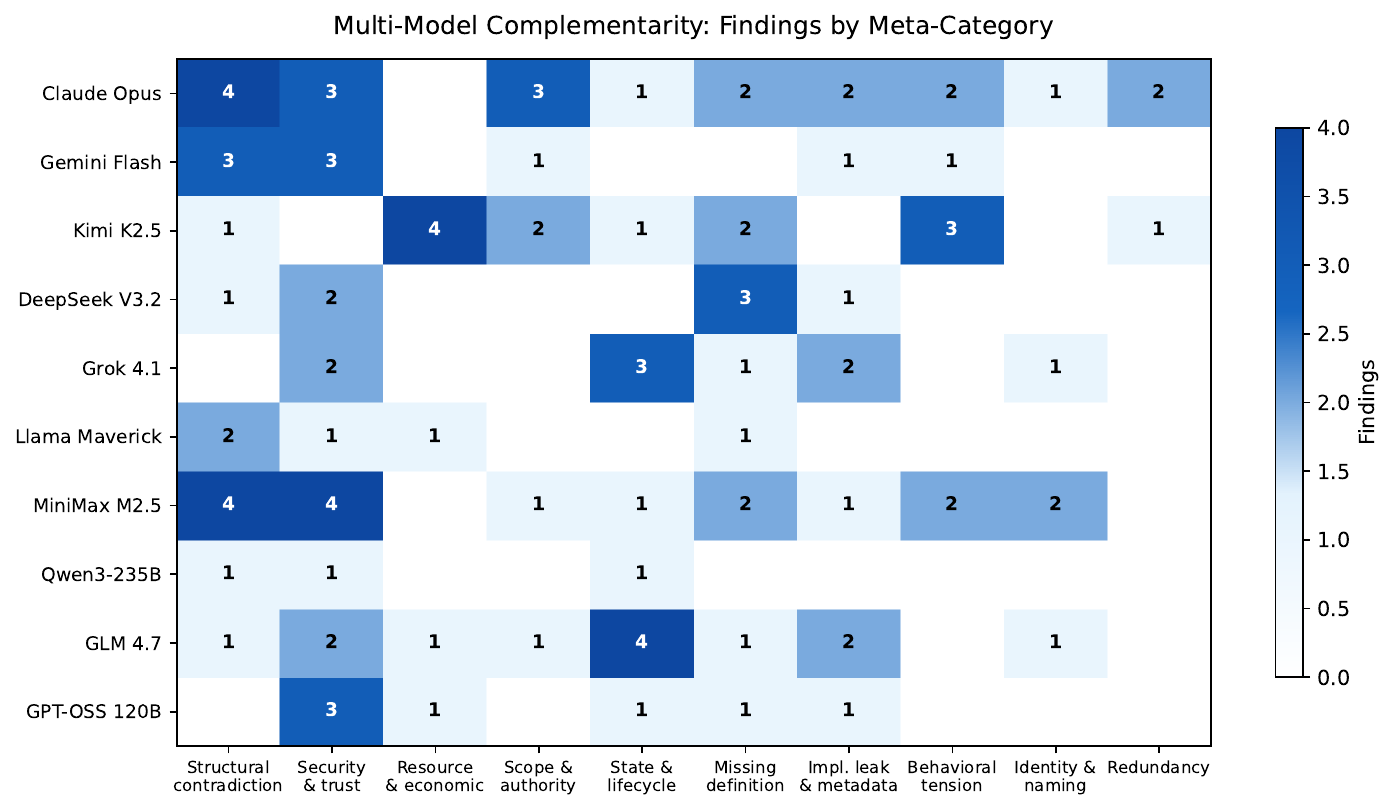}
\caption{Multi-model complementarity heat map for Claude Code. Findings
  are clustered into ten meta-categories. Security \& trust is the only
  category found by 9/10 models; Resource \& economic findings come
  almost exclusively from Kimi~K2.5; MiniMax~M2.5 contributes the
  broadest coverage (8/10 categories). The sparsity pattern demonstrates
  that models are complementary, not redundant.}
\label{fig:heatmap}
\end{figure}

The category explosion in Claude Code---107 unique categories for 116
findings---quantifies this: each model invents its own taxonomy. The
categories don't converge. The coverage does.
The ten meta-categories in \Cref{fig:heatmap} are analyst-curated for
interpretability, and alternative groupings are possible.

\subsection{Convergence Properties}
\label{sec:convergence}

\begin{figure}[H]
\centering
\includegraphics[width=0.85\textwidth]{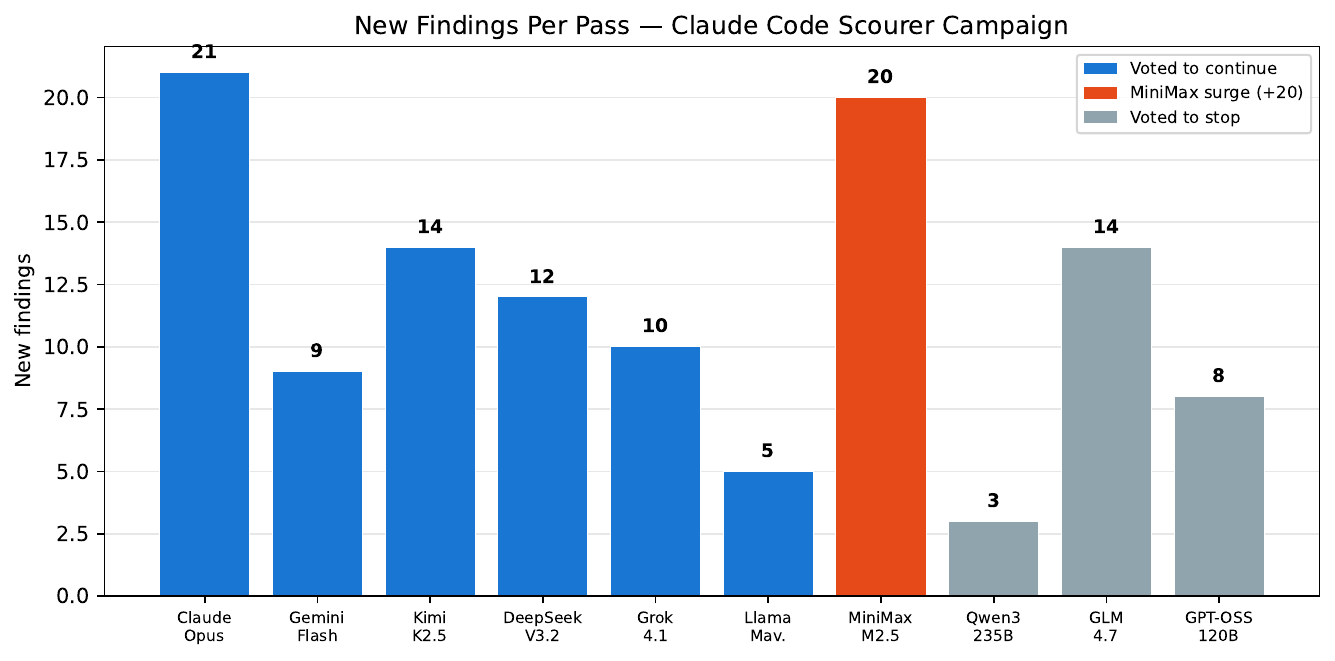}
\caption{New findings per scourer pass for Claude Code. The MiniMax
  M2.5 surge at pass~7 (+20 findings after pass~6 produced only~5)
  demonstrates that specific models bring viewpoints capable of
  reopening exploration even after apparent convergence. Gray bars
  indicate passes that voted to stop; the stopping criterion requires
  three consecutive ``no'' votes.}
\label{fig:per-pass}
\end{figure}

The non-monotonic convergence of Claude Code is notable. Pass~7's
surge of 20 findings (after pass~6 produced~5) suggests that specific
models bring viewpoints capable of reopening exploration even after
apparent convergence. The three-consecutive-``no'' criterion addresses
this for the Claude Code campaign. The Codex and Gemini campaigns, each
stopped at a single ``no,'' could have missed such a surge.

\subsection{Structural Profiles}
\label{sec:structural-profiles}

The prompt AST parser (\Cref{sec:ast}) provides quantitative structural
profiles for each vendor's prompt, moving the architectural taxonomy
from qualitative description to measurement.

\begin{table}[H]
\centering
\caption{Structural profiles from AST analysis.}
\label{tab:structural-profiles}
\begin{tabular}{@{}lrrrr@{}}
\toprule
Metric & \multicolumn{2}{c}{Claude Code} & Codex & Gemini \\
\cmidrule(lr){2-3}
 & v2.1.50 & v2.1.71 & CLI & CLI \\
\midrule
Nodes & 80 & 110 & 183 & 151 \\
Max depth & 2 & 4 & 5 & 4 \\
Sections & 0 & 12 & 16 & 21 \\
Top-level directives & 7 & 2 & 0 & 0 \\
Unclassified nodes & 23 (29\%) & 1 (1\%) & 13 (7\%) & 1 (1\%) \\
\bottomrule
\end{tabular}
\end{table}

\begin{figure}[H]
\centering
\begin{tikzpicture}
\begin{axis}[
    ybar stacked,
    width=0.85\textwidth,
    height=6cm,
    bar width=20pt,
    xlabel={},
    ylabel={Percentage of nodes},
    ymin=0, ymax=105,
    symbolic x coords={Claude v2.1.50, Claude v2.1.71, Codex CLI, Gemini CLI},
    xtick=data,
    x tick label style={font=\small},
    legend style={at={(1.02,1)}, anchor=north west, font=\small},
    grid=major,
    grid style={gray!15},
]
\addplot[fill=notable!80] coordinates
    {(Claude v2.1.50,61) (Claude v2.1.71,45) (Codex CLI,75) (Gemini CLI,72)};
\addplot[fill=alarming!80] coordinates
    {(Claude v2.1.50,35) (Claude v2.1.71,13) (Codex CLI,9) (Gemini CLI,5)};
\addplot[fill=curious!80] coordinates
    {(Claude v2.1.50,4) (Claude v2.1.71,16) (Codex CLI,15) (Gemini CLI,22)};
\addplot[fill=concerning!80] coordinates
    {(Claude v2.1.50,0) (Claude v2.1.71,25) (Codex CLI,0) (Gemini CLI,1)};
\legend{Behavior, Tool schema, Environment, Memory}
\end{axis}
\end{tikzpicture}
\caption{Channel distribution across vendor prompts. Claude Code v2.1.71
  is the only prompt with a substantial memory channel (25\%). Codex and
  Gemini CLI are behavioral-dominant ($>$70\%). The version evolution from
  v2.1.50 to v2.1.71 shows tool definitions migrating out of the prompt
  text into API parameters.}
\label{fig:channels}
\end{figure}
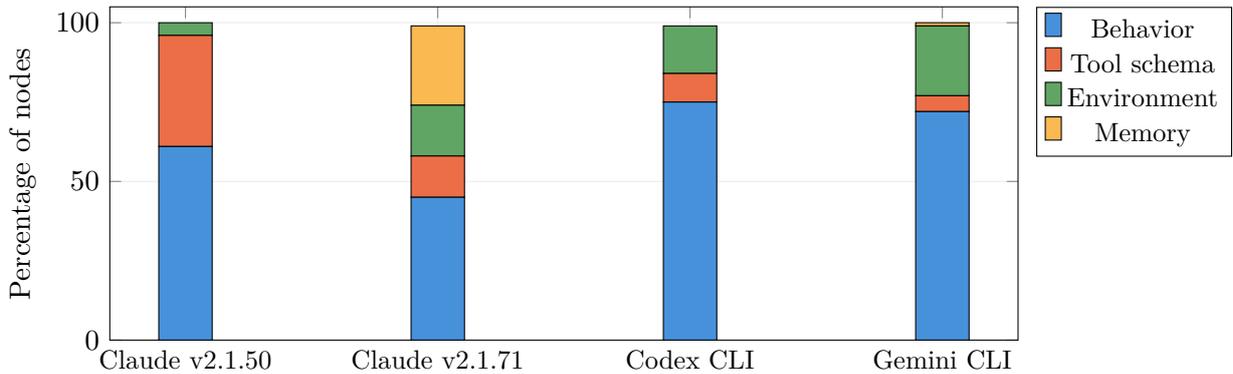

\subsubsection{Version Evolution}

The AST differ applied to two versions of Claude Code (v2.1.50 and
v2.1.71) reveals a systematic architectural evolution:

\begin{itemize}[leftmargin=*]
  \item \textbf{Flat blob $\to$ sectioned document.} Depth increased
    from 2 to 4; sections went from 0 to 12. The prompt gained
    hierarchical structure.
  \item \textbf{Prose $\to$ structured lists.} Paragraph count
    decreased while list count increased---behavioral constraints moved
    from running text into enumerated items.
  \item \textbf{Tool extraction.} Tool definitions previously embedded
    in the prompt text were extracted to the API \texttt{tools}
    parameter, reducing tool\_schema channel from 35\% to 13\%.
  \item \textbf{Memory channel emergence.} An entire \texttt{auto
    memory} section (28 nodes, 25\% of the prompt) was added---a new
    functional channel with no precedent in the earlier version.
  \item \textbf{Classification clarity.} Unclassified nodes dropped
    from 29\% to 1\%. Sections make semantic role assignment possible
    where flat structure left it ambiguous.
\end{itemize}

Analysis of 49 sessions captured via instrumented proxy (March 3--7,
2026) revealed that \textbf{system prompt structure is invariant within
a version.} All per-session variation was confined to the Environment
section (dynamic injection of working directory, git status, date).
Behavioral, safety, tool, memory, and format channels never changed
between sessions---only between version updates. This has a practical
implication: AST analysis can be cached per-version, with only the
environment section requiring per-session re-parsing.

\section{Discussion}
\label{sec:discussion}

\subsection{Prompts as Software Artifacts}

The analogy between system prompts and software systems is not
metaphorical. The same structural taxonomy (monolith, modular, flat)
applies to both, and it predicts the same classes of failure:
monoliths accumulate contradictions at internal subsystem boundaries;
modular systems produce bugs at composition seams; flat systems trade
capability for consistency.

This is not a coincidence. It is Conway's
Law~\citep{conway1968committees} applied to a new medium: the structure
of the prompt reflects the structure of the team that produced it.
Claude Code's monolithic prompt---with its TodoWrite subsystem
contradicting its commit workflow---reads like the output of separate
teams adding capabilities to a shared artifact without cross-team
integration testing.

The implication is that system prompts need the same engineering
infrastructure that conventional software has: linters for internal
consistency, tests for behavioral contracts, CI/CD pipelines that
catch regressions when one section is edited. Arbiter's directed
evaluation rules are a prototype of this infrastructure.

\subsection{The Observer's Paradox}

The thesis---\textbf{the agent that resolves the conflict cannot be the
agent that detects it}---has a precise formulation: the heuristic that
enables an LLM to navigate contradictory instructions is the same
heuristic that prevents it from recognizing those instructions as
contradictory. Detection requires a different vantage point.

The executing LLM smooths over contradictions via ``judgment.'' When
Claude Code encounters ``ALWAYS use TodoWrite'' and ``NEVER use
TodoWrite'' in the same context, it picks one. The system
works---most of the time. This is precisely why the contradictions
persist: they cause no visible errors, only invisible inconsistency.

\subsection{Severity and Epistemic Confidence}

The directed and undirected phases use different severity scales, and
this difference is informative.

\textbf{Directed analysis} uses an impact scale: \emph{critical}
(structurally guaranteed to cause incorrect behavior), \emph{major}
(will cause problems under specific conditions), \emph{minor}
(maintenance concern).

\textbf{Undirected scouring} uses an epistemic confidence scale:
\emph{curious} (pattern noticed), \emph{notable} (worth
investigating), \emph{concerning} (likely problematic), \emph{alarming}
(structurally guaranteed to fail).

These scales are orthogonal. A finding can be epistemically ``curious''
but operationally critical, or epistemically ``alarming'' but
operationally irrelevant. The multi-model scouring provides epistemic
provenance---which model found which finding---enabling downstream
consumers to weight findings by the reliability of their source.

\subsection{Translation Boundary: Epistemic vs. Normative Truth}
\label{sec:translation-boundary}

Recent neutrosophic work on LLM uncertainty suggests that scalar outputs
alone can collapse meaningful distinctions, and that tensorized outputs
(including structured declared losses) can recover them. We treat this
as a \emph{representation transfer}, not a domain identity claim.

Epistemic uncertainty asks what is true, indeterminate, or false about a
proposition. Arbiter's adjudication layer asks what is admissible,
ambiguous, or conflicting under tiered policy constraints. These are
different truth domains. The transfer we make is structural: richer
output channels can preserve distinctions that scalar channels collapse.

For follow-on adjudication experiments, we pre-specify an operational
persistent-indeterminacy criterion to avoid unfalsifiable ``just one more
pass'' loops: a case is persistent-high-I when $I \ge \tau_I$ after at
least $N_{\min}$ passes, with marginal change $|\Delta I| < \epsilon$
for $k$ consecutive passes, and no allowed canon/context mutation lowers
$I$ below $\tau_I$.

\subsection{Cost}
\label{sec:cost}

The total cost of analyzing all three vendor prompts was \$0.27 USD,
verified against OpenRouter billing records:

\begin{table}[H]
\centering
\caption{API cost by vendor.}
\label{tab:cost-vendor}
\begin{tabular}{@{}lrr@{}}
\toprule
Vendor & Passes & Actual Cost \\
\midrule
Claude Code & 10 & \$0.236 \\
Codex CLI & 2 & \$0.012 \\
Gemini CLI & 3 & \$0.014 \\
\midrule
\textbf{Total} & \textbf{15} & \textbf{\$0.263} \\
\bottomrule
\end{tabular}
\end{table}

The cost per finding across all three analyses is \$0.002. At US
federal minimum wage (\$7.25/hour), the entire cross-vendor campaign
costs less than three minutes of human labor. This is not a
methodological footnote---it is a result. It means that system prompt
analysis at this level of thoroughness is accessible to any developer
with API access, not just teams with dedicated security budgets.

\begin{figure}[H]
\centering
\includegraphics[width=0.75\textwidth]{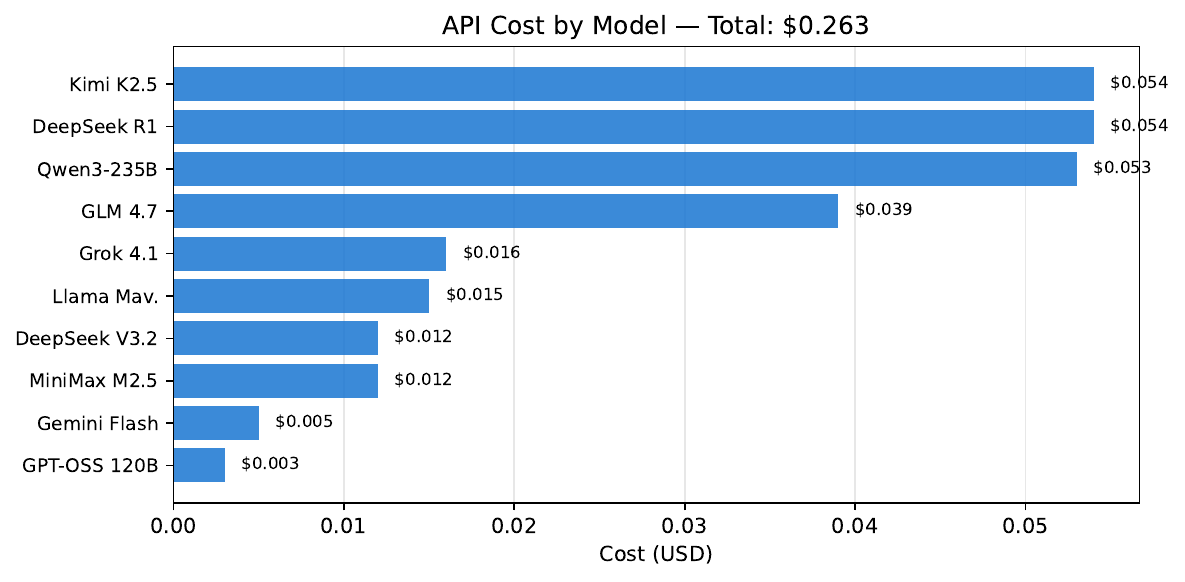}
\caption{API cost by model. The three most expensive models (Kimi~K2.5,
  DeepSeek~R1, Qwen3-235B) account for 61\% of total cost, driven by
  retries and reasoning token overhead. GPT-OSS~120B provided 8
  findings for \$0.003.}
\label{fig:cost}
\end{figure}

\subsection{Limitations}

\textbf{Static analysis only.} This work analyzes system prompt text,
not runtime behavior. A contradiction in the prompt may or may not
manifest as incorrect behavior, depending on how the executing LLM
resolves it. Recent work by~\citet{gloaguen2026evaluating} provides
partial runtime evidence from an adjacent domain: repository-level
context files (AGENTS.md, CLAUDE.md) that contain the same kinds of
natural-language instructions we analyze. They find that LLM-generated
context files \emph{reduce} task success rates while increasing
inference cost by 20--23\%, and that ``unnecessary requirements make
tasks harder''---consistent with our finding that instruction
redundancy and contradiction create measurable interference. Their
reasoning token data (14--22\% increase with context files) suggests
that the structural conditions we identify via static analysis
manifest as cognitive load at runtime.

\textbf{Scourer findings are LLM-generated.} The undirected scourer
asks LLMs to analyze prompts, and LLMs can fabricate observations.
Version~1 claimed two mitigations, and neither holds. Multi-model
execution does not confirm findings across models, because each pass
is instructed not to repeat prior findings; the protocol forbids the
confirmation that was claimed. And no human review was recorded: the
author read the findings while assembling the taxonomy, but no
per-finding verdict was written down, so no precision figure can be
reported for the 152. Recording blind per-finding verdicts, with a
second labeler, is the first item of follow-on work.

\textbf{Shared-model blind spots.} Multi-model agreement is not proof of
independence. Frontier models may share training-distribution blind
spots, creating false consensus. This motivates explicit adversarial
challenge rounds (including human spoiler roles) in follow-on studies.

\textbf{Directed analysis limited to one vendor.} The 56-block
decomposition and 21-pattern ground truth exist only for Claude Code.
The 21 patterns are also both the ground truth for the directed
evaluator and the reported directed-analysis result; no precision or
recall of the evaluator against an independent labeling is reported,
and a second labeler has not been used.

\textbf{One prompt per architecture class.} The architecture-to-failure-
class correlation rests on one prompt per class, and the classes differ
in size (a five-fold range in lines) and in campaign effort (ten
scourer passes for Claude Code against two and three for the others,
and directed analysis on Claude Code only). The correlation is a
hypothesis to be tested on further prompts per class, not an
established relationship.

\textbf{Gemini CLI prompt is composed.} Unlike Claude Code and Codex
CLI, whose prompts were captured as-is, Gemini CLI's prompt was
composed from TypeScript render functions with all feature flags
enabled.

\textbf{Upstream context processing effects.} This paper analyzes frozen
prompt artifacts. In deployed systems, context compaction, summarization,
or demand paging can alter clause visibility and therefore change
interference detectability. Integration-level evaluation under processed
context is future work.

\textbf{Cost data excludes subscription allocation.} Claude Opus 4.6
(one pass) was billed via Anthropic subscription. The reported \$0.27
is a lower bound.

\textbf{Three vendors.} The architectural taxonomy is grounded in three
examples. The architecture--failure-mode claim should be read as a
strong correlation in this sample, not a universal law. Additional
vendors may reveal architectures that don't fit this taxonomy.

\subsection{Responsible Disclosure}

Two findings have potential user impact:

\textbf{Gemini CLI: save\_memory data loss.} The structural guarantee
that saved preferences are deleted during history compression affects
real users who rely on the memory feature for long sessions. Google
filed and patched a related symptom (issue \#16213, PR \#16914) without
addressing the schema-level root cause. The schema bug remains in the
publicly available source code as of February 2026.

\textbf{Claude Code: recursive agent spawning.} The Task tool's
\texttt{Tools: *} specification includes the Task tool itself, enabling
unbounded recursive agent spawning. This is a known architectural
choice (depth is controlled by other mechanisms) but the system prompt
does not make this explicit.

Both findings derive from publicly available artifacts.

\section{Conclusion}
\label{sec:conclusion}

System prompts are the least-tested, most-consequential software
artifacts in modern AI systems. They govern the behavior of agents that
write code, manage files, execute shell commands, and interact with
external services. Yet they receive no static analysis, no integration
testing, and no cross-team review for internal consistency.

Three architectural patterns---monolithic, flat, modular---produce
three characteristic classes of failure. This taxonomy transfers from
conventional software engineering because the underlying structural
dynamics are the same: growth produces boundary contradictions,
modularity produces seam defects, simplicity buys consistency at the
cost of capability.

Successive models in a map-passing chain surfaced different categories
of findings. Whether that is a property of the models or of the
instruction to avoid repetition is the first question this version
leaves open; the same-model control that answers it costs under a
dollar.

The total cost of comprehensive cross-vendor analysis is twenty-seven
cents---less than three minutes of minimum wage labor, less than a
single finding from a human security audit. The tools exist. The data
is clear. Nobody is checking.

\section{Code and Data Availability}
\label{sec:availability}

Code, analysis scripts, and reproducibility artifacts are available at
\url{https://github.com/<org-or-user>/arbiter} (paper snapshot:
tag \texttt{v0.1.1}, commit \texttt{f7b2a56}).

A frozen archival snapshot is available via Zenodo DOI:
\url{https://doi.org/10.5281/zenodo.18929834}.

The deterministic reproduction workflow and artifact manifest checks are
documented in \texttt{ARTIFACT.md} in the repository.

\bibliography{references}

@article{wei2022chain,
  title={Chain-of-Thought Prompting Elicits Reasoning in Large Language Models},
  author={Wei, Jason and Wang, Xuezhi and Schuurmans, Dale and Bosma, Maarten and Ichter, Brian and Xia, Fei and Chi, Ed and Le, Quoc V and Zhou, Denny},
  journal={Advances in Neural Information Processing Systems},
  volume={35},
  pages={24824--24837},
  year={2022}
}

@article{yao2023tree,
  title={Tree of thoughts: Deliberate problem solving with large language models},
  author={Yao, Shunyu and Yu, Dian and Zhao, Jeffrey and Shafran, Izhak and Griffiths, Tom and Cao, Yuan and Narasimhan, Karthik},
  journal={Advances in neural information processing systems},
  volume={36},
  pages={11809--11822},
  year={2023}
}

@inproceedings{schulhoff2023ignore,
  title={Ignore this title and HackAPrompt: Exposing systemic vulnerabilities of LLMs through a global prompt hacking competition},
  author={Schulhoff, Sander and Pinto, Jeremy and Khan, Anaum and Bouchard, Louis-Fran{\c{c}}ois and Si, Chenglei and Anati, Svetlina and Tagliabue, Valen and Kost, Anson and Carnahan, Christopher and Boyd-Graber, Jordan},
  booktitle={Proceedings of the 2023 Conference on Empirical Methods in Natural Language Processing},
  pages={4945--4977},
  year={2023}
}

@article{greshake2023not,
  title={Not What You've Signed Up For: Compromising Real-World {LLM}-Integrated Applications with Indirect Prompt Injection},
  author={Greshake, Kai and Abdelnabi, Sahar and Mishra, Shailesh and Endres, Christoph and Holz, Thorsten and Fritz, Mario},
  journal={Proceedings of the 16th ACM Workshop on Artificial Intelligence and Security},
  pages={79--90},
  year={2023}
}

@article{conway1968committees,
  title={How Do Committees Invent?},
  author={Conway, Melvin E},
  journal={Datamation},
  volume={14},
  number={4},
  pages={28--31},
  year={1968}
}

@book{newman2021building,
  title={Building microservices: designing fine-grained systems},
  author={Newman, Sam},
  year={2021},
  publisher={" O'Reilly Media, Inc."}
}

@article{zheng2023judging,
  title={Judging llm-as-a-judge with mt-bench and chatbot arena},
  author={Zheng, Lianmin and Chiang, Wei-Lin and Sheng, Ying and Zhuang, Siyuan and Wu, Zhanghao and Zhuang, Yonghao and Lin, Zi and Li, Zhuohan and Li, Dacheng and Xing, Eric and others},
  journal={Advances in neural information processing systems},
  volume={36},
  pages={46595--46623},
  year={2023}
}

@article{liu2023prompt,
  title={Prompt injection attack against llm-integrated applications},
  author={Liu, Yi and Deng, Gelei and Li, Yuekang and Wang, Kailong and Wang, Zihao and Wang, Xiaofeng and Zhang, Tianwei and Liu, Yepang and Wang, Haoyu and Zheng, Yan and others},
  journal={arXiv preprint arXiv:2306.05499},
  year={2023}
}

@article{wang2022self,
  title={Self-consistency improves chain of thought reasoning in language models},
  author={Wang, Xuezhi and Wei, Jason and Schuurmans, Dale and Le, Quoc and Chi, Ed and Narang, Sharan and Chowdhery, Aakanksha and Zhou, Denny},
  journal={arXiv preprint arXiv:2203.11171},
  year={2022}
}

@article{parnas1972criteria,
  title={On the criteria to be used in decomposing systems into modules},
  author={Parnas, David Lorge},
  journal={Communications of the ACM},
  volume={15},
  number={12},
  pages={1053--1058},
  year={1972},
  publisher={ACm New York, NY, USA}
}

@article{gloaguen2026evaluating,
  title={Evaluating AGENTS. md: Are Repository-Level Context Files Helpful for Coding Agents?},
  author={Gloaguen, Thibaud and M{\"u}ndler, Niels and M{\"u}ller, Mark and Raychev, Veselin and Vechev, Martin},
  journal={arXiv preprint arXiv:2602.11988},
  year={2026}
}

\appendix

\section{Scourer Convergence Data}
\label{app:convergence}

\begin{table}[H]
\centering
\caption{Claude Code v2.1.50 scourer passes.}
\label{tab:convergence-claude}
\begin{tabular}{@{}rlrrl@{}}
\toprule
Pass & Model & New & Cumulative & Continue? \\
\midrule
1 & Claude Opus 4.6 & 21 & 21 & yes \\
2 & Gemini 2.0 Flash & 9 & 30 & yes \\
3 & Kimi K2.5 & 14 & 44 & yes \\
4 & DeepSeek V3.2 & 12 & 56 & yes \\
5 & Grok 4.1 & 10 & 66 & yes \\
6 & Llama 4 Maverick & 5 & 71 & yes \\
7 & MiniMax M2.5 & 20 & 91 & yes \\
8 & Qwen3-235B & 3 & 94 & no \\
9 & GLM 4.7 & 14 & 108 & no \\
10 & GPT-OSS 120B & 8 & 116 & no \\
\bottomrule
\end{tabular}
\end{table}

\begin{table}[H]
\centering
\caption{Codex CLI scourer passes.}
\begin{tabular}{@{}rlrrl@{}}
\toprule
Pass & Model & New & Cumulative & Continue? \\
\midrule
1 & DeepSeek V3.2 & 10 & 10 & yes \\
2 & Grok 4.1 & 5 & 15 & no \\
\bottomrule
\end{tabular}
\end{table}

\begin{table}[H]
\centering
\caption{Gemini CLI scourer passes.}
\begin{tabular}{@{}rlrrl@{}}
\toprule
Pass & Model & New & Cumulative & Continue? \\
\midrule
1 & DeepSeek V3.2 & 12 & 12 & yes \\
2 & Qwen3-235B & 5 & 17 & yes \\
3 & GLM 4.7 & 4 & 21 & no \\
\bottomrule
\end{tabular}
\end{table}

\section{Scourer Prompt Templates}
\label{app:prompts}

\subsection{First Pass}

\begin{quote}
\ttfamily\small
You are exploring a system prompt. Not auditing it, not checking it
against rules---just reading it carefully and noting what you find
interesting.

\medskip

``Interesting'' is deliberately vague. Trust your judgment. You might
notice: instructions that seem to contradict each other; rules stated
multiple times in different places; implicit assumptions that aren't
declared; surprising structural choices; scope ambiguities; things that
would confuse a model trying to follow all instructions simultaneously;
interactions between distant parts of the prompt; anything else that
catches your attention.

\medskip

After documenting what you found, document what you DIDN'T explore.
What areas did you skim? What questions occurred to you that you didn't
pursue?

\medskip

Finally: should we send another explorer after you? Would another pass,
armed with your map, find things you missed?
\end{quote}

\subsection{Subsequent Passes}

\begin{quote}
\ttfamily\small
You are exploring a system prompt. Previous explorers have already been
through it and left you their map. Your job is to go where they didn't.

\medskip

DO NOT repeat their findings. They found what they found. You are
looking for what they missed, what they flagged as unexplored, and
anything their framing caused them to overlook.

\medskip

[Previous findings and unexplored territory injected here]

\medskip

Be honest about diminishing returns. Set should\_send\_another to FALSE
if: most of your findings are refinements or restatements of existing
ones; the unexplored territory is mostly about runtime behavior; you
found fewer than 3 genuinely new findings; the prior passes have
already covered the major structural, security, operational, and
semantic categories.

\medskip

It is better to say ``enough'' than to pad findings.
\end{quote}

\section{Directed Analysis: Claude Code Interference Patterns}
\label{app:interference}

\begin{table}[H]
\centering
\caption{21 hand-labeled interference patterns in Claude Code v2.1.50.}
\label{tab:interference-full}
\small
\begin{tabular}{@{}rlllc@{}}
\toprule
\# & Type & Blocks & Severity & Static? \\
\midrule
1 & Direct contradiction & TodoWrite mandate $\leftrightarrow$ Commit restriction & Critical & Yes \\
2 & Direct contradiction & TodoWrite reinforcement $\leftrightarrow$ Commit restriction & Critical & Yes \\
3 & Direct contradiction & TodoWrite mandate $\leftrightarrow$ PR restriction & Critical & Yes \\
4 & Direct contradiction & TodoWrite reinforcement $\leftrightarrow$ PR restriction & Critical & Yes \\
5 & Scope overlap & TodoWrite mandate $\leftrightarrow$ TodoWrite tool & Major & Yes \\
6 & Priority ambiguity & Security policy $\leftrightarrow$ Security policy (dup) & Minor & Yes \\
7 & Scope overlap & Conciseness $\leftrightarrow$ TodoWrite (overhead) & Major & No \\
8 & Scope overlap & Conciseness $\leftrightarrow$ WebSearch (sources) & Minor & Yes \\
9 & Scope overlap & Task tool search $\leftrightarrow$ Explore agent & Major & Yes \\
10 & Scope overlap & Read-before-edit (general) $\leftrightarrow$ Edit tool & Minor & Yes \\
11 & Scope overlap & Read-before-edit (general) $\leftrightarrow$ Write tool & Minor & Yes \\
12 & Scope overlap & No-new-files (tone) $\leftrightarrow$ Write tool & Minor & Yes \\
13 & Scope overlap & No-new-files (tone) $\leftrightarrow$ Edit tool & Minor & Yes \\
14 & Scope overlap & Dedicated tools (policy) $\leftrightarrow$ Bash tool & Minor & Yes \\
15 & Scope overlap & Dedicated tools (policy) $\leftrightarrow$ Grep tool & Minor & Yes \\
16 & Scope overlap & No time estimates $\leftrightarrow$ Asking questions & Minor & Yes \\
17 & Implicit dependency & Commit restrictions $\leftrightarrow$ General Bash policy & Minor & Yes \\
18 & Implicit dependency & Plan mode tool restrictions $\leftrightarrow$ Tool policy & Minor & Yes \\
19 & Scope overlap & No-emoji (tone) $\leftrightarrow$ Edit tool & Minor & Yes \\
20 & Scope overlap & No-emoji (tone) $\leftrightarrow$ Write tool & Minor & Yes \\
21 & Priority ambiguity & Parallel calls $\leftrightarrow$ Commit workflow ordering & Minor & Yes \\
\bottomrule
\end{tabular}
\end{table}

\section{Cost Breakdown by Model}
\label{app:cost}

\begin{table}[H]
\centering
\caption{Per-model API costs from OpenRouter billing records.}
\label{tab:cost-model}
\begin{tabular}{@{}lrr@{}}
\toprule
Model & Calls & Actual Total \\
\midrule
Kimi K2.5 & 2 & \$0.054 \\
DeepSeek R1 & 1 & \$0.054 \\
Qwen3-235B & 3 & \$0.053 \\
GLM 4.7 & 2 & \$0.039 \\
Grok 4.1 Fast & 5 & \$0.016 \\
Llama 4 Maverick & 3 & \$0.015 \\
DeepSeek V3.2 & 3 & \$0.012 \\
MiniMax M2.5 & 1 & \$0.012 \\
Gemini 2.0 Flash & 2 & \$0.005 \\
GPT-OSS 120B & 4 & \$0.003 \\
\midrule
\textbf{Total} & \textbf{26} & \textbf{\$0.263} \\
\bottomrule
\end{tabular}
\end{table}

Call count exceeds pass count (15) due to retries (Kimi K2.5 output
length limit), intermediate experiments (DeepSeek R1), and growing
prompt size as accumulated findings are passed forward. The Qwen3-235B
total includes one anomalous 175K-token prompt (\$0.046 alone), likely
a routing artifact. Cost per finding: \$0.002.

\end{document}